\documentstyle[aps,prd,preprint]{revtex}
\begin{document}
\draft

\title {\bf Electrostatic in Reissner-Nordstr\"om space-time with a \\conical
defect}
\author{J. Spinelly and V. B. Bezerra } 
\address{Departamento de F\'{\i}sica, Universidade Federal da Para\'{\i}ba\\
Caixa Postal 5008, 58051-970 Jo\~{a}o Pessoa,PB,Brazil\\
\parbox{14 cm}{\medskip\rm\indent
We calculate the electrostatic potential generated by a point charge at rest
in Reissner-Nordstr\"om space-time with a conical defect. An expression for 
the self-energy is also presented.}}
\maketitle
$      $

It is well known that the gravitational field modifies the electrostatic
interaction of a charged particle in such a way that the particle 
experiences a finite self-force\cite{smith,leaute}. The origin of this 
force comes from the space-time curvature associated with the 
gravitational field. On the other hand, even in the absence of 
curvature, such as in the conical 
space-time of an infinite straight cosmic string\cite{vilenkin}, it 
was shown that a charged point particle\cite{linet1} or a linear charge 
distribution\cite{mello} placed at rest in this background becomes 
subject to a finite 
repulsive electrostatic self-force. In this case, the origin of this force 
is the distortion in the  particle field caused by the lack of global 
flatness of the space-time of a cosmic string. Therefore, the modifications
of the electrostatic potential comes from two contributions, one of geometric
origin and the other of topological one. 

Some authors have suggested that the most simple exact solutions of Einstein's
equations can easily be generalized to include  a conical 
defect\cite{shellard}. Such space-times have been considered in different
context and some investigations were done in these backgrounds\cite{marcelo}. 

In this paper we determine the expression for the electrostatic potential
generated by a point charge held stationary in the space-time of Reissner-
Nordstr\"om pierced by a cosmic string and also determine the 
self-energy. These results extend previous one obtained 
by Linet\cite{linet2} in the case of a Schwarzschild background with conical
defect.

The Reissner-Nordstr\"om space-time endowed with a conical defect takes the
following form 
\begin{small}
\begin{eqnarray}
ds^{2}=&&-\left( 1-\frac{2E}{br}+\frac{q^{2}}{b^{2}r^{2}}\right) dt^{2}+
\left( 1-\frac{2E}{br}+\frac{q^{2}}{b^{2}r^{2}}\right) ^{-1}dr^{2}
\nonumber\\
&&+r^{2}d\theta^{2}+b^{2}r^{2}\sin ^{2}\theta d\varphi ^{2},  \label{eq1}
\end{eqnarray}
\end{small}
where $b$ is a parameter quantifying the conical defect. Assuming that the
conical defect is associated with a cosmic string, then the parameter $b$ 
is given by $b= 1 - 4\mu$, where $\mu$ is the linear mass density of the 
cosmic string. The quantity $E$ is 
the energy of the black hole\cite{aryal} and $q$ is the electric charge. 
Space-times with a conical defect are geometrically constructed by removing 
a wedge, that is, by requiring that the azimuthal angle around the axis 
runs over the range $0< \phi < 2\pi b$. In the present case, in addition 
to the geometric procedure, we have to correct the mass term with the factor 
$\frac {1}{b}$ due to the fact that in the presence of a string, the energy 
at infinity and the Schwarzschild mass parameter are not identical
\cite{aryal}. The same correction must be applied to the electrostatic term 
due to Gauss theorem. In the calculations that follow we will consider the 
region of space-time outside the outer horizon, that is, for $r>r_{+}$, where
$r_{+}=\left( E + \sqrt{E^{2}-q^{2}}\right) /b$, and assume that
$E^{2}>q^{2}.$

Consider a point test charge $e$, held stationary at 
$\left( r_{o},\theta_{o},\varphi _{o}\right)$, with $r_{o}>r_{+}$. 
In this case the current density $J^{i}$ vanishes and the charge density
$J^{0}$ is given by

\begin{equation}
J^0=\frac e{br^2\sin \theta }\delta \left( r-r_o\right) \delta \left( \theta
-\theta _o\right) \delta \left( \varphi -\varphi _o\right) .  \label{2}
\end{equation}

Now, let us take Maxwell's equations in the space-time given by metric
(\ref{eq1}) with source being a point test charge $e$. In the present 
case we have just one equation for the component $A_{0}$ which is given by

\begin{eqnarray}
\partial_{i}\left( \sqrt{-g}F^{i0}\right) = 4\pi \sqrt{-g}J^0,  \label{3}
\end{eqnarray}
where $F_{i0}= \partial _i{A_0}$.

Therefore, in the space-time under consideration, the equation for ${A_0}$ can
be written as 
\begin{eqnarray}
\frac{\partial }{\partial r}\left(r^2 \frac{\partial A_{0}}{\partial r}
 \right) &+&\left(1-\frac{2E}{br}+\frac{q^2}{b^2 r^2} \right)^{-1}
   \left[\frac{1}{\sin \theta }\frac{\partial }{\partial\theta }
     \left(\sin\theta \frac{\partial A_{0}}{\partial \theta } \right)
   +\frac{1}{b^2 \sin^2 \theta}\frac{\partial^{2}A_{0}}{\partial \varphi^{2}}
       \right]\nonumber\\
     &=& -\frac{4\pi e}{b\sin \theta }\delta \left( r-r_{o} \right) 
       \delta \left(\theta -\theta _{o} \right) \delta \left( \varphi 
          -\varphi _{o} \right) .
\label{4}
\end{eqnarray}

Doing the following substitutions\cite{linet4}

\begin{equation}
r=r_{s}+r_{-}  \label{5}
\end{equation}

\begin{equation}
A_{0}=\frac{r-r_{-}}{r}A_{0}^{s},  \label{6}
\end{equation}
where $r_{-}=E/b-M$ and $M^{2}=\left( E^{2}-q^{2}\right) /b^{2}$,
eq.(\ref{4}) turns into

\begin{eqnarray}
\frac{\partial }{\partial r_{s}}\left( r_{s}^{2}\frac{\partial A_{0}^{s}}{
\partial r_{s}}\right) &+&\left( 1-\frac{2M}{r_{s}}\right) ^{-1}\left[ \frac{1
}{\sin \theta }\frac{\partial }{\partial \theta }\left( \sin \theta \frac{
\partial A_{0}^{s}}{\partial \theta }\right) 
+\frac{1}{b^{2}\sin ^{2}\theta }
\frac{\partial ^{2}A_{0}^{s}}{\partial \varphi ^{2}}\right] \nonumber \\
&=&-\frac{4\pi e}{b\sin \theta }\frac{r_{s}}{r_{s}+r_{-}}\delta \left(
r_{s}-r_{so}\right) \delta \left( \theta -\theta _{o}\right) \delta \left(
\varphi -\varphi _{o}\right) ,  \label{7}
\end{eqnarray}
which can be rewritten as

\begin{eqnarray}
\frac{\partial }{\partial r_{s}}\left( r_{s}^{2}\frac{\partial A_{0}^{s}}{%
\partial r_{s}}\right) &+&\left( 1-\frac{2M}{r_{s}}\right) ^{-1}\left[ \frac{1%
}{\sin \theta }\frac{\partial }{\partial \theta }\left( \sin \theta \frac{%
\partial A_{0}^{s}}{\partial \theta }\right) +\frac{1}{b^{2}\sin ^{2}\theta }%
\frac{\partial ^{2}A_{0}^{s}}{\partial \varphi ^{2}}\right] \nonumber\\
&=&-\frac{4\pi e^{\prime }}{b\sin \theta }\delta \left( r_{s}-r_{so}\right)
\delta \left( \theta -\theta _{o}\right) \delta \left( \varphi -\varphi
_{o}\right) ,  \label{8}
\end{eqnarray}
where $e^{\prime }=er_{so}/\left( r_{so}+r_{-}\right) $. Equation (%
\ref{8}) is formally identical to the partial differential equation 
for the electrostatic potential in the space-time of Schwarzschild
with a conical defect\cite{linet2}. Proceeding in analogy with
Linet\cite{linet2} and assuming that $\varphi_{o}=\pi $ and $1/2<b<1$, 
we find that the solution for (\ref{8}) is given by

\begin{equation}
A_{0}^{s}\left( r_{s},\theta ,\varphi \right) =V_{s}^{\ast }\left(
r_{s},\theta ,\varphi \right) +V_{s}^{b}\left( r_{s},\theta ,\varphi \right)
+\frac{e^{\prime }M}{br_{so}r_{s}},  \label{9}
\end{equation}
where

\begin{equation}
V_{s}^{\ast }\left( r_{s},\theta ,\varphi \right) =\left\{ 
\begin{array}{c}
V_{cs}\left[ r_{s},\sigma _{0}\left( \theta ,\varphi \right) \right] +V_{cs}%
\left[ r_{s},\sigma _{1}\left( \theta ,\varphi \right) \right] \\ 
V_{cs}\left[ r_{s},\sigma _{0}\left( \theta ,\varphi \right) \right] \\ 
V_{cs}\left[ r_{s},\sigma _{0}\left( \theta ,\varphi \right) \right] +V_{cs}%
\left[ r_{s},\sigma _{-1}\left( \theta ,\varphi \right) \right]
\end{array}\right. 
\begin{array}{c}

, \\ 

, \\ 

,

\end{array}
\begin{array}{c}
0<\varphi <\pi /b-\pi \\ 
\pi /b-\pi <\varphi <3\pi -\pi /b \\ 
3\pi -\pi /b<\varphi <2\pi
\end{array}.  \label{10}
\end{equation}

Note that solution given by eq.(\ref{9}) is fixed taking into account the 
requirement that charge $q$ of the black hole and charge $e$ of the test
particle contribute to the potential.

The term $V_{cs}$ is the solution of Copson\cite{copson}, which is a solution
of eq.(\ref{8}) for $b=1$ and whose expression is\cite{linet3}

\begin{equation}
V_{cs}\left[ r_{s},\sigma \right] =\frac{e^{\prime }}{r_{so}r_{s}}\frac{%
\left( r_{s}-M\right) \left( r_{so}-M\right) -M^{2}\sigma }{\left[ \left(
r_{s}-M\right) ^{2}+\left( r_{so}-M\right) ^{2}-M^{2}-2\left( r_{s}-M\right)
\left( r_{so}-M\right) \sigma +M^{2}\sigma ^{2}\right] ^{1/2}}  \label{11}
\end{equation}
with

\begin{eqnarray}
\sigma _{0}\left( \theta ,\varphi \right) &=&\cos \theta \cos \theta
_{o}+\sin \theta \sin \theta _{o}\cos [ b\left( \varphi -\pi \right)]  
\label{12} \\
\sigma _{1}\left( \theta ,\varphi \right) &=&\cos \theta \cos \theta
_{o}+\sin \theta \sin \theta _{o}\cos [ b\left( \varphi +\pi \right)]  
\label{13} \\
\sigma _{-1}\left( \theta ,\varphi \right) &=&\cos \theta \cos \theta
_{o}+\sin \theta \sin \theta _{o}\cos [ b\left( \varphi -3\pi \right)] .  
\label{14}
\end{eqnarray}

The potential $V_{s}^{b}$ that appears in eq.(\ref{9}) is 
given by\cite{linet2}

\begin{equation}
V_{s}^{b}\left( r_{s},\theta ,\varphi \right) =\frac{1}{2\pi b}%
\int_{0}^{\infty }V_{cs}\left[ r_{s},k\left( \theta ,x\right) \right]
F_{b}\left( \varphi ,x\right) dx,  \label{15}
\end{equation}
where $k\left( \theta ,x\right) $ and $F_{b}\left( \varphi
,x\right)$ are 

\begin{equation}
k\left( \theta ,x\right) =\cos \theta \cos \theta _{o}-\sin \theta \sin
\theta _{o}\cosh x  \label{16}
\end{equation}
and

\begin{equation}
F_{b}\left( \varphi ,x\right) =-\frac{\sin \left( \varphi -\pi /b\right) }{%
\cosh \left( x/b\right) +\cos \left( \varphi -\pi /b\right) }+\frac{\sin
\left( \varphi +\pi /b\right) }{\cosh \left( x/b\right) +\cos \left( \varphi
+\pi /b\right) }.  \label{17}
\end{equation}

Now, using eqs. (\ref{5}) and (\ref{6}), we can rewrite the 
electrostatic potential given by eq.(\ref{9}) as

\begin{equation}
A_{0}=V^{\ast }\left( r,\theta ,\varphi \right) +V^{b}\left( r,\theta
,\varphi \right) +\frac{eM}{br_{o}r},  \label{18}
\end{equation}
where

\begin{equation}
V^{\ast }\left( r,\theta ,\varphi \right) =\left\{ 
\begin{array}{c}
V_{c}\left[ r,\sigma _{0}\left( \theta ,\varphi \right) \right] +V_{c}\left[
r,\sigma _{1}\left( \theta ,\varphi \right) \right] \\ 
V_{c}\left[ r,\sigma _{0}\left( \theta ,\varphi \right) \right] \\ 
V_{c}\left[ r,\sigma _{0}\left( \theta ,\varphi \right) \right] +V_{c}\left[
r,\sigma _{-1}\left( \theta ,\varphi \right) \right]
\end{array}
\right. 
\begin{array}{c}, \\ , \\ ,
\end{array}
\begin{array}{c}
0<\varphi <\pi /b-\pi \\ 
\pi /b-\pi <\varphi <3\pi -\pi /b \\ 
3\pi -\pi /b<\varphi <2\pi
\end{array},  \label{19}
\end{equation}
and $V_{c}$ 

\begin{equation}
V_{c}\left[ r,\sigma \right] =\frac{e}{r_{o}r}\frac{\left( r-E/b\right)
\left( r_{o}-E/b\right) -M^{2}\sigma }{\left[ \left( r-E/b\right)
^{2}+\left( r_{o}-E/b\right) ^{2}-M^{2}-2\left( r-E/b\right) \left(
r_{o}-E/b\right) \sigma +M^{2}\sigma ^{2}\right] ^{1/2}}.  \label{20}
 \end{equation}

The potential $V^{b}$ obeys the following relation

\begin{equation}
V^{b}\left( r,\theta ,\varphi \right) =\frac{1}{2\pi b}\int_{0}^{\infty
}V_{c}\left[ r,k\left( \theta ,x\right) \right] F_{b}\left( \varphi
,x\right) dx.  \label{21}
\end{equation}

Doing the same procedure as in\cite{linet2} we can verify that eq.(\ref{9})
is a solution of eq.(\ref{8}), or returning to the original radial coordinate,
we conclude that eq.(\ref{18}) is a solution of eq.(\ref{4}), that is, 
eq.(\ref{18}) is the electrostatic potential generated by a point charge 
$e$ placed at rest in the space-time of a charged black hole
(Reissner-Nordstr\"{o}m space-time) with a cosmic string passing through it.

Now, let us consider the electrostatic potential eq.(\ref{18}) in a
neighborhood of the point $\left( r_{o},\theta _{o}, \pi \right) $ in the
region defined by $ \frac{\pi}{b} < \phi < 3\pi - \frac{\pi}{b}$. The terms
$V^{b}\left(r,\theta \right) $ and $eM/brr_{o}$ are finite at the position of
the test charge. The term  $V^{\ast }\left( r,\theta \right) $ 
is infinite at that point and corresponds to the Coulombian potential.
Therefore the electrostatic self-force comes from $V^{reg} \equiv
V^{b}\left( r,\theta \right) +eM/brr_{o}$. From the definition of energy for
an arbitrary charge distribution we get that the self-energy is given by

\begin{equation}
U_{self}\left( r_{o},\theta _{o}\right) =\frac{1}{2}\left( eV^{b}\left(
r_{o},\theta _{o}\right) +\frac{e^{2}M}{br_{o}^{2}}\right) .  \label{22}
\end{equation}

Putting eq. (\ref{21}) into eq.(\ref{22}) and using eq.(\ref{17}), we get
the following result

\begin{equation}
U_{self}\left( r_{o},\theta _{o}\right) =\frac{e^{2}M}{2br_{o}^{2}}-\frac{%
e\sin \left( \pi /b\right) }{2\pi }\int_{0}^{\infty }V_{c}\left[
r_{o},k\left( \theta _{o},x\right) \right] \frac{dx}{\cosh \left( x/b\right)
-\cos \left( \pi /b\right) }  \label{23}
\end{equation}

Equation (\ref{23}) is the gravitationally induced electrostatic 
self-energy on a test point particle at rest in the space-time of 
Reissner-Nordstr\"{o}m with a conical defect(cosmic string). It is important
to call attention to the fact that this result brings together the geometric
and topological aspects. The geometric contribution is connected with the 
curvature of the space-time and the topological features are due to the lack
of spherical symmetry produced by the conical defect. 

From the present results we can get the particular cases which has already
been obtained corresponding to the electrostatic self-force in the 
Schwarzschild black hole with\cite{linet2} and without\cite{leaute} a conical
defect and in the cosmic string space-time\cite{linet1}.

\newpage
\noindent
{\bf Acknowledgment}\\
\noindent
This work was partially supported by CNPq and CAPES. We are grateful to
Prof. B. Linet for useful discussions.

\end{document}